\documentclass[epj]{webofc}
\usepackage[utf8]{inputenc}
\usepackage[varg]{txfonts}   
\usepackage{booktabs}
\usepackage{xcolor}
\definecolor{darkred}{rgb}{0.4,0.0,0.0}
\definecolor{darkgreen}{rgb}{0.0,0.4,0.0}
\definecolor{darkblue}{rgb}{0.0,0.0,0.4}
\usepackage[bookmarks,linktocpage,colorlinks,
    linkcolor = darkred,
    urlcolor  = darkblue,
    citecolor = darkgreen]{hyperref}
\usepackage{subfigure}
\wocname{EPJ Web of Conferences}
\woctitle{Lattice2017}
\begin{document}

\begin{flushright}
KEK-TH-2009
\end{flushright}

\vspace{-0.3cm}

%
\selectlanguage{english}
\title{%
Complex Langevin simulation of QCD at finite density \\
and low temperature using the deformation technique
}
\author{%
\firstname{Keitro} \lastname{Nagata}\inst{1,2}\and 
\firstname{Jun} \lastname{Nishimura}\inst{2,3} \and
\firstname{Shinji}  \lastname{Shimasaki}\inst{2,4}\fnsep\thanks{Speaker,
\email{shinji.shimasaki@keio.jp}}
}
\institute{%
Center of Medical Information Science, Kochi Medical School, Kochi
University
\and
KEK Theory Center, High Energy Accelerator Research Organization, Tsukuba 305-0801, Japan
\and
Department of Particle and Nuclear Physics, 
School of High Energy Accelerator Science,  \\
 Graduate University for Advanced Studies (SOKENDAI), Tsukuba 305-0801, Japan
\and
Research and Education Center for Natural Sciences, \\
 Keio University, Hiyoshi 4-1-1, Yokohama, Kanagawa 223-8521, Japan
}
\abstract{%
We study QCD at finite density and low temperature 
by using the complex Langevin method. 
We employ the gauge cooling to control the unitarity norm and 
introduce a deformation parameter in the Dirac operator 
to avoid the singular-drift problem. 
The reliability of the obtained results are judged 
by the probability distribution of the magnitude of the drift term. 
By making extrapolations
with respect to the deformation parameter
using only the reliable results,
we obtain results for the original system. 
We perform simulations on a $4^3\times8$ lattice and 
show that our method works well even in the region 
where the reweighting method fails due to the severe sign problem. 
As a result we observe a delayed onset of the baryon number 
density as compared with the phase-quenched model, 
which is a clear sign of the Silver Blaze phenomenon. 
  }
\maketitle
\section{Introduction}\label{intro}

Investigating the QCD phase diagram at finite density 
and temperature is one of the central issues 
in modern theoretical physics.
In particular, QCD at high density and low temperature
is interesting due to its relevance to the state of
dense matter that is considered to be realized in astronomical
objects like neutron stars.
It is known, however, that QCD in this parameter region 
suffers from a severe sign problem, which makes it
practically inaccessible by
conventional lattice QCD simulations based 
on the importance sampling.

One promising approach to the sign problem
that has attracted much attention in recent years 
is the complex Langevin method (CLM) \cite{Parisi:1984cs,Klauder:1983sp}.
The CLM is based on the stochastic time evolution 
for the complexified dynamical variables using
the complex Langevin equation.
Since this method does not rely on the probabilistic interpretation 
of the Boltzmann weight in the path integral formulation 
of the original theory,
one has a chance to solve the sign problem.
It is known, however, that the equivalence 
between the CLM and the original path integral does not always 
hold \cite{Aarts:2009uq,Aarts:2011ax}.
In order to prove this equivalence,
we actually need to satisfy the condition that 
the probability distribution of 
the magnitude of drift term decays 
exponentially or faster \cite{Nagata:2016vkn}. 
If this condition is met, one can show that the CLM 
gives the correct result.
If the drift distribution falls off 
more slowly than exponential,
the proof of the equivalence does not go through,
and the results obtained by the CLM can be simply wrong.
Thus, in applying the CLM, it is extremely important to
judge the reliability of the obtained results
by monitoring the asymptotic behavior of the drift distribution.

In fact, it is known that there are two causes
for the slow decay of the drift distribution.
One is the excursion problem, where
the complexified dynamical variables run deeply into the imaginary direction.
The other is the singular-drift problem, 
where the probability distribution of the configurations 
does not vanish at the singularity of the drift term.
In order to overcome these problems and to 
enlarge the applicability region of the CLM,
various techniques have been 
developed \cite{Seiler:2012wz, Aarts:2016qhx,Ito:2016efb,Bloch:2017ods}.
For instance, 
the gauge cooling \cite{Seiler:2012wz}
is a technique for avoiding the excursion problem 
by making use of 
the complexified gauge transformation.
(It can be used also to avoid the 
singular-drift problem \cite{Nagata:2016alq}.)
In order to avoid the singular-drift problem,
one can use the deformation technique \cite{Ito:2016efb}.
In the case of QCD, for instance, 
one can deform the Dirac operator by adding a deformation parameter
in such a way that the eigenvalue distribution does not cover the origin. 
The result of the original theory can then be recovered 
by extrapolating the deformation parameter to zero
using only the data points that pass the reliability test.

Here we show for the first time that
the CLM can be used to investigate
QCD at high density and low temperature.
We perform simulations on a $4^3\times 8$ lattice and 
calculate the baryon number density as a function of 
the quark chemical potential. (See also ref.~\cite{Sinclair:2016nbg} for related work.)
We use the gauge cooling for the excursion problem 
and the deformation technique for the singular-drift problem.
To test the reliability of the obtained results, 
we monitor the probability distribution of the drift term.
We find, in particular, 
that the onset of the baryon number density occurs
at smaller chemical potential than 
in the phase-quenched model,
which is a clear sign of
the so-called Silver Blaze phenomenon \cite{Cohen:2003kd}.

The rest of this article is organized as follows.
In Sec.~\ref{sec-2}, we first
review the application of the CLM to lattice QCD at finite density. 
We then explain the condition required for the validity of the CLM
as well as the techniques we use to meet this condition
such as the gauge cooling and the deformation technique.
In Sec.~\ref{sec-3}, we present our results 
obtained on a $4^3\times 8$ lattice.
Sec.~\ref{sec-4} is devoted to a summary and discussions.

\section{Method}\label{sec-2}


\subsection{Application of the CLM to lattice QCD}\label{sec-2-1}

We consider lattice QCD defined on a four-dimensional lattice 
of the size $N_t$ and $N_s$ 
in the temporal and spatial directions, respectively.
(The lattice spacing is set to unity throughout this article.)
The partition function is given by
\begin{align}
  Z = \int \prod_{x\mu}dU_{x\mu} \,
  \mathrm{det}M(U,\mu) \, e^{-S_{\rm g}(U)} \ ,
  \label{Z}
\end{align}
where $x=(x_1,x_2,x_3,x_4)$ labels the sites on the lattice,
and $U_{x\mu}\in \mathrm{SU}(3)$ ($\mu=1,2,3,4$)
are the link variables.
The gauge action $S_{\rm g}(U)$ is given by
\begin{align}
  S_{\rm g}
= -\frac{\beta}{6}
\sum_{x}\sum_{\mu>\nu}\mathrm{tr}[U_{x,\mu\nu}+U_{x,\mu\nu}^{-1}] \ ,
\end{align}
where 
$U_{x,\mu\nu}=U_{x\mu}U_{x+\hat\mu,\nu}U_{x+\hat\nu,\mu}^{-1}U_{x\nu}^{-1}$ 
with $\hat\mu$ being the unit vector in the $\mu$ direction.
The fermion determinant
$\mathrm{det}M(U,\mu)$ in \eqref{Z}
is defined with the fermion matrix for 
the staggered fermion with four flavors given by
\begin{align}
  M(U,\mu)_{xy}=m\delta_{xy}+\sum_{\nu=1}^{4}\frac{\eta_\nu(x)}{2}
\left(e^{\mu \delta_{\nu 4}}U_{x\nu}\delta_{x+\hat\nu , y}
    -e^{-\mu \delta_{\nu 4}}U_{x-\hat\nu , \nu}^{-1}
   \delta_{x-\hat\nu , y}\right) \ ,
    \label{fermion matrix}
  \end{align}
where $m$ is the degenerate quark mass and 
$\mu$ is the quark chemical potential.
As usual, we introduce 
the sign factors $\eta_\mu=(-1)^{x_1+\cdots+x_{\mu-1}}$,
which play the role of the gamma matrices.
Note that $\epsilon_x M(U,\mu)_{xy}\epsilon_y=M(U,-\mu^*)^\dagger$
holds,
where $\epsilon_x=(-1)^{x_1+x_2+x_3+x_4}$ plays the role of $\gamma_5$.
This identity suggests for $\mu\neq 0$
that $\mathrm{det}M(U,\mu)$ can be complex and the sign problem occurs.

Here we investigate the lattice QCD \eqref{Z} at $\mu\neq 0$
using the CLM, in which 
the link variables
$U_{x\mu}\in \mathrm{SU}(3)$ are complexified 
to $\mathcal U_{x\mu}\in \mathrm{SL}(3,\mathbb{C})$.
The discretized version of the complex Langevin equation is given by
\begin{align}
  \mathcal U^{(\eta)}_{x\mu}(t+\epsilon)
  =\exp\left(i\sum_{a=1}^{8}\lambda_a
\left[-\epsilon v_{ax\mu}(\mathcal U^{(\eta)}(t))
    +\sqrt{\epsilon}\eta_{ax\mu}(t)\right]  \right) \ 
\mathcal U^{(\eta)}_{x\mu}(t) \ ,
    	\label{CLM}
\end{align}
where $t$ is the discretized time with the step size $\epsilon$, 
$\lambda_a \ (a=1,\cdots,8)$ are the 
SU(3) generators normalized by 
$\mathrm{tr}(\lambda_a \lambda_b) = \delta_{ab}$
and
$\eta_{ax\mu}(t)$ is the real Gaussian noise
with the distribution $\exp\{-\frac{1}{4}\eta_{ax\mu}(t)^2\}$.
The so-called drift term $v_{ax\mu}(\mathcal U)$ is 
defined as the holomorphic extension of the one
\begin{align}
  v_{ax\mu}(U)
=\left.\frac{d}{dz}S(e^{iz\lambda_a}U_{x\mu})\right|_{z=0} 
  \label{drift}
  \end{align}
for the unitary link variables $U_{x\mu}$,
where $S(U)=S_g(U)- \mathrm{log} \, \mathrm{det} M(U,\mu)$.

The CLM enables us to calculate the expectation values of 
observables that admit the holomorphic extension 
by taking the ensemble average with thermalized configurations 
obtained by solving eq.~\eqref{CLM}.
As a typical observable, we consider the baryon number density
\begin{align}
  n
  &=\frac{1}{3N_V}\frac{\partial}{\partial \mu}\log Z \nonumber\\
  &=\frac{1}{3N_V}\left\langle 
\sum_{x} \frac{\eta_4(x)}{2}\mathrm{tr}
  (e^{\mu }M^{-1}_{x+\hat4, x}U_{x4}
+e^{-\mu }M^{-1}_{x-\hat4, x}U_{x-\hat4,4}^{-1})\right\rangle \ ,
  \label{baryon}
\end{align}
where $N_V=N_tN_s^3$ and the symbol $\langle\cdots\rangle$ represents 
the expectation value with respect to \eqref{Z}.

\subsection{Condition for the validity of the CLM}

It is known that the results obtained by the CLM 
are not always correct \cite{Aarts:2009uq,Aarts:2011ax},
and therefore one has to judge whether they are reliable or not.
For that purpose,
we need to investigate 
how the probability distribution 
of the drift term behaves asymptotically 
at large magnitude \cite{Nagata:2016vkn}.
It is shown under certain assumptions that 
the CLM gives the correct result if the distribution
decays exponentially or faster.
Roughly speaking, frequent appearance of a large drift
invalidates the equivalence between the CLM 
and the path integral of the original theory.
Note that this problem cannot be solved by making the step size smaller
or by using the adaptive step size.
In this work, we use the magnitude of the drift term defined by
\begin{align}
\widetilde{\mathcal{U}}
=\sqrt{\frac{1}{8N_V} \sum_{x \mu} \sum_{a=1}^{8}
\left| v_{ax\mu}(\mathcal{U})\right|^2} \ ,
\end{align}
and monitor its probability distribution for thermalized configurations. 

There are actually two causes for
the frequent appearance of a large drift, 
which makes the CLM fail.
One is the excursion problem,
which occurs when the complexified link variables 
$\mathcal U_{x\mu}\in \mathrm{SL}(3,\mathbb{C})$
go far away from $\mathrm{SU}(3)$ \cite{Aarts:2009uq,Aarts:2011ax}.
The other is the singular-drift problem, 
which occurs when the fermion matrix has
near-zero eigenvalues \cite{Nishimura:2015pba}.
In order to avoid these problems and to make the CLM work, 
one needs to modify the complex Langevin dynamics in some way or another.

\subsection{Gauge cooling for the excursion problem}

For the excursion problem, 
we use the gauge cooling \cite{Seiler:2012wz},
which amounts to making a complexified gauge transformation 
after each Langevin update
so that the complexified link variables become 
as close to $\mathrm{SU}(3)$ as possible.
The gauge transformation is determined by minimizing the unitarity norm
\begin{align}
    \mathcal N
  =\frac{1}{4N_V}\sum_{x\mu}\mathrm{tr}
  \left[(\mathcal U_{x\mu})^\dagger 
\mathcal U_{x\mu}+(\mathcal U_{x\mu}^{-1})^\dagger \mathcal U^{-1}_{x\mu}
    -2\times \mathbf{1}_{3\times 3}\right] \ ,
  \end{align}
which measures the distance of the complexified link variables 
$\mathcal U_{x\mu}\in \mathrm{SL}(3,\mathbb{C})$ from $\mathrm{SU}(3)$.
While the gauge cooling changes the Langevin dynamics nontrivially, 
the changes do not affect the proof of the equivalence between the CLM 
and the path integral of the original theory
due to the covariance of the drift term and the invariance of the 
observable under the complexified gauge transformation \cite{Nagata:2015uga}.
Therefore, the proof goes through
if the same condition for reliability is satisfied.
Indeed this technique played a crucial role
in applying the CLM to 
finite density QCD at high temperature \cite{Sexty:2013ica}.

\subsection{Deformation for the singular-drift problem}
\label{sec-2-2}

In the high density and low temperature region,
we have to deal also with the singular-drift problem.
For that purpose,
we use the deformation technique \cite{Ito:2016efb},
which amounts to deforming the fermion matrix 
by introducing a parameter so that the singular-drift problem is avoided. 
In this work, we consider the deformation
\begin{align}
  M(U,\mu)_{xy}\to M(U,\mu)_{xy}+i\alpha \eta_{4}(x)\delta_{xy} \ ,
  \end{align}
where $M(U,\mu)_{xy}$ is the fermion matrix
defined in \eqref{fermion matrix} and 
$\alpha$ is the deformation parameter.
Note that this deformation corresponds formally 
to adding an imaginary chemical potential in the continuum theory,
which implies that
the chiral symmetry and the Lorentz symmetry are kept intact.
For a sufficiently large $\alpha$, 
the eigenvalue spectrum of the fermion matrix 
develops a gap in the imaginary direction, 
and thus the singular-drift problem can be avoided.
Using the symmetry under $\alpha \leftrightarrow -\alpha$,
we can fit the data points to a polynomial in $\alpha^2$ and 
obtain results for the original theory by 
making $\alpha\to 0$ extrapolations
using only the data points that pass the reliability test.

\section{Results}\label{sec-3}

In this section, we show our results for finite density QCD
with $\beta=5.7$, $m=0.05$
and various values of the quark chemical potential up to 
$\mu=0.7$ on a $4^3\times 8$ lattice.
In applying the CLM,
we use a fixed step size $\epsilon=10^{-4}$
and the total Langevin time $50\sim 150$.
The gauge cooling and the deformation technique are used to
avoid the excursion problem and the singular-drift problem, 
respectively.
We calculate the baryon number density \eqref{baryon} for the
deformed theory,
and make extrapolations to the zero deformation parameter
using only the reliable data points.


\begin{figure}
  \centering
  \includegraphics[width=6cm]{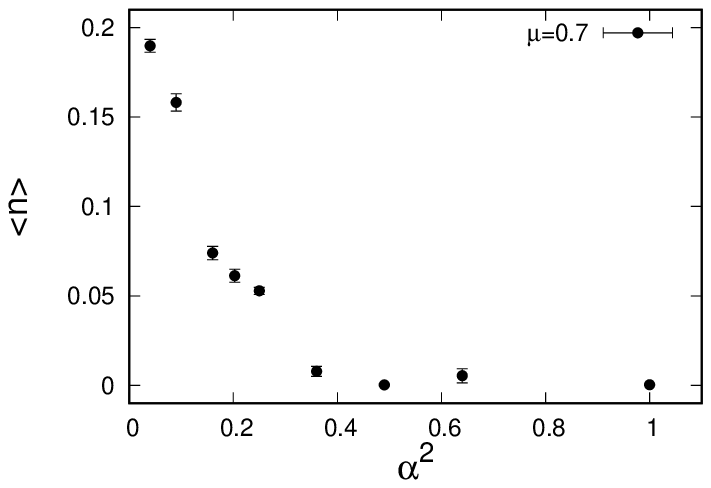}
  \hspace{5mm}
  \includegraphics[width=6cm]{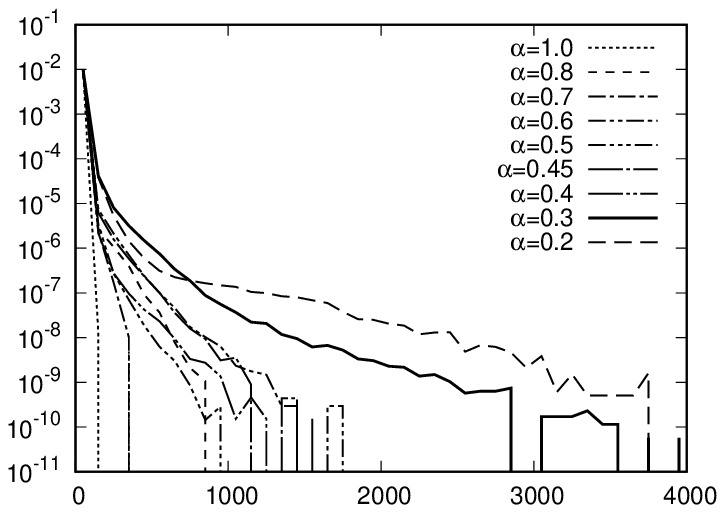}
  \\[5mm]
  \includegraphics[width=6cm]{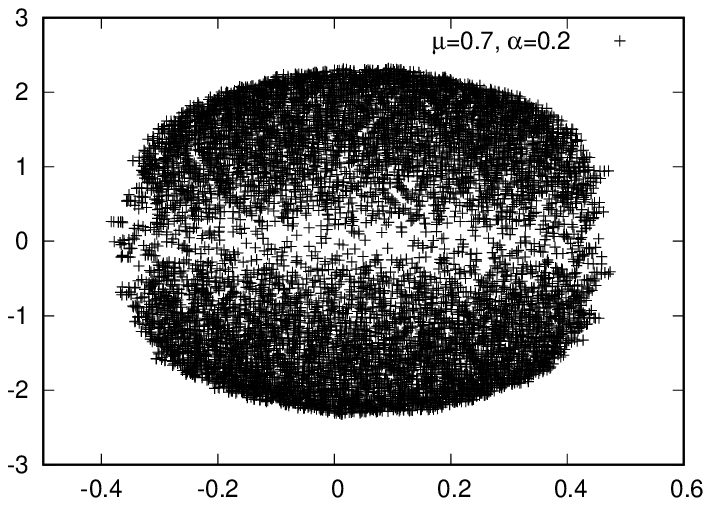}
  \hspace{5mm}
  \includegraphics[width=6cm]{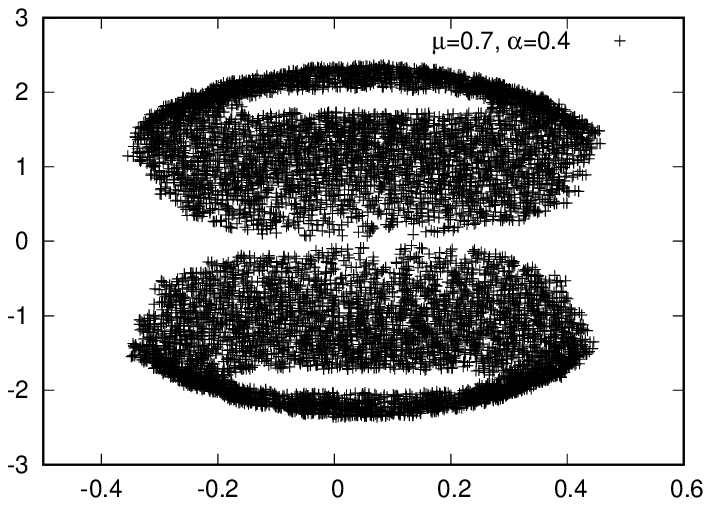}
 \caption{
Results obtained at $\mu=0.7$.
(Top-Left) The baryon number density is plotted against $\alpha^2$, 
where $\alpha$ is the deformation parameter.
(Top-Right) The probability distribution of the drift term
is shown in a semi-log plot for various $\alpha$.
The distribution for $\alpha=0.2$ and $0.3$ 
has a power-law tail, which implies that
the corresponding data points in the left figure
are not reliable.
(Bottom-Left) The eigenvalue distribution of 
the fermion matrix is shown for $\alpha=0.2$. 
The distribution covers the origin, 
which causes the power-law tail in the 
probability distribution of the drift term.
(Bottom-Right) The eigenvalue distribution of the fermion matrix 
is shown for $\alpha=0.4$. There exists a gap near the origin, which
helps in avoiding large drifts.}
\label{n small mu=0.7}
\end{figure}

Let us first explain how we choose the data points 
to be used for the extrapolation
by taking the $\mu=0.7$ case as an example.
In figure \ref{n small mu=0.7} (Top-Left), we plot
the baryon number density 
against $\alpha^2$.

First we have to choose the data points 
that pass the reliability test.
For that, we look at
the probability distribution of the drift term 
shown in figure \ref{n small mu=0.7} (Top-Right),
where each line corresponds to each data point in figure
\ref{n small mu=0.7} (Top-Left).
We find that the results for $\alpha=0.2, 0.3$ are not reliable 
because the probability distribution of the drift term falls off
with a power law.

The behavior of the drift distribution
in figure \ref{n small mu=0.7} (Top-Right) 
is consistent with
the eigenvalue distribution of the fermion matrix
shown 
in figure \ref{n small mu=0.7} (Bottom-Left) and (Bottom-Right)
for $\alpha=0.2$ and $\alpha=0.4$, respectively. 
For $\alpha=0.2$, we find that
the eigenvalue distribution covers the origin, 
which causes the frequent appearance of a large drift term.
For $\alpha=0.4$, on the other hand,
we observe no eigenvalues close to the origin due to the gap made 
by the relatively large $\alpha$,
which avoids the appearance of a large drift term.

\begin{figure}
  \centering
  \includegraphics[width=12cm]{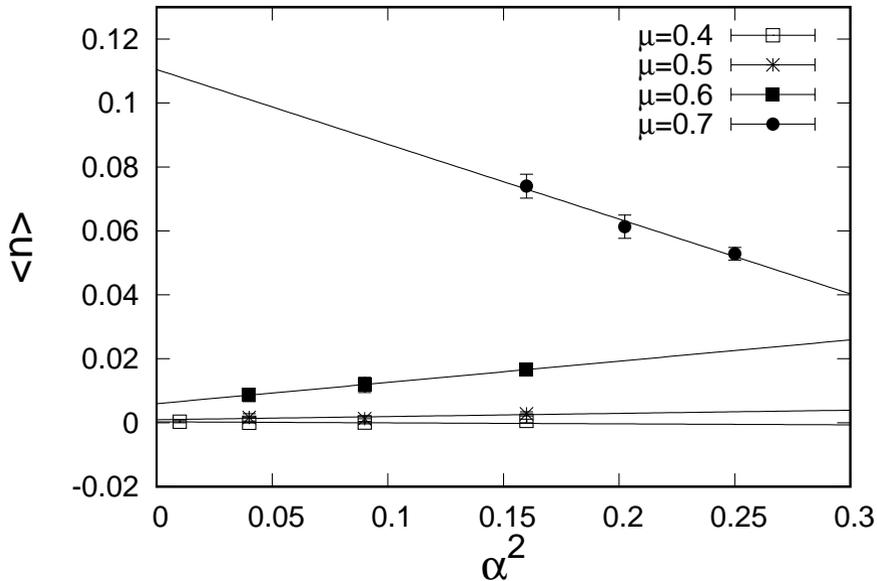} 
  \caption{
The baryon number density is plotted for $\mu=0.4, 0.5, 0.6, 0.7$
against $\alpha^2$, where $\alpha$ is the deformation parameter.}
\label{n small mu=all}
  \end{figure}

While the results for $\alpha\gtrsim 0.4$ are reliable, 
we cannot use the data points at too large $\alpha$,
which should
include higher order corrections with respect to $\alpha^2$.
Note, in particular, that
the baryon number density seems to vanish identically for $\alpha\geq 0.6$,
which suggests that there is a phase transition at $\alpha \sim 0.6$.
Hence, for $\mu=0.7$, we use the data points obtained with
$\alpha=0.4, 0.45, 0.5$ and make a linear extrapolation
as is shown in figure \ref{n small mu=all}.
We do the same thing for other values of $\mu$,
and plot the reliable data points in the same figure.
The linear extrapolation to $\alpha=0$ is made 
using $\alpha=0.2, 0.3, 0.4$ for $\mu=0.5, 0.6$
and $\alpha=0.1, 0.2, 0.3$ for $\mu=0.4$.

Let us add that the deformation technique is useful 
not only in avoiding the singular-drift problem, 
but also in stabilizing the simulations.
In fact, at $\mu=0.4$, the singular-drift problem may not occur
even without the deformation \cite{Nagata:2016mmh}.
However, the history of observables have large spikes,
which makes it difficult to reduce the statistical errors
within reasonable amount of statistics.
This problem is solved by the deformation even with 
$\alpha$ as small as $0.1$,
which clearly suggests another virtue of 
using the deformation in the CLM.

\begin{figure}
  \centering
  \includegraphics[width=12cm]{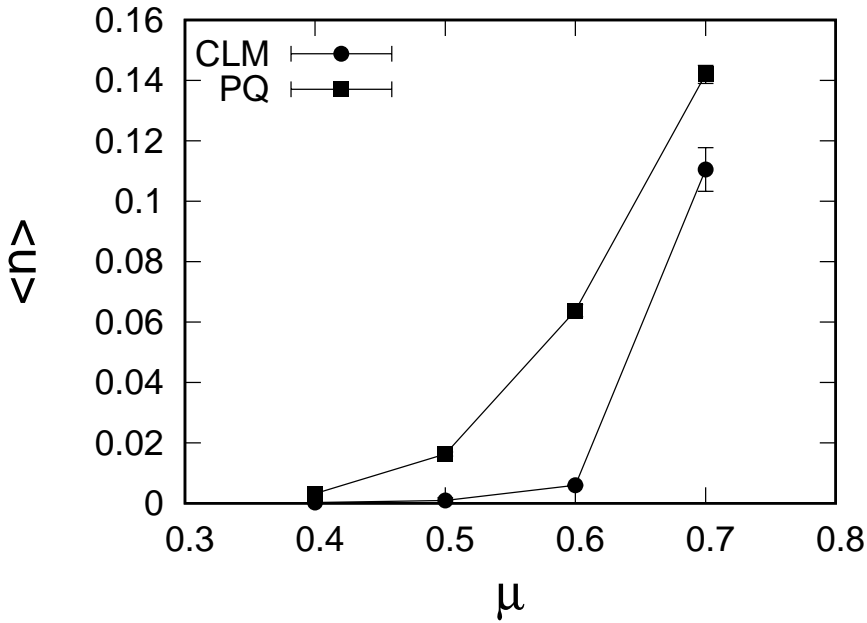} 
  \caption{
The baryon number density is plotted against $\mu$.
The circles represent the results obtained by the CLM
using the $\alpha\to 0$ extrapolation shown 
in figure \ref{n small mu=all}.
The squares represent the results 
obtained by the phase-quenched model.}
\label{n small mu=all r}
  \end{figure}

In figure \ref{n small mu=all r},
the circles represent
the baryon number density in the original theory 
obtained by the $\alpha\to 0$ extrapolations shown 
in figure \ref{n small mu=all}.
We find that the baryon number density is zero for $\mu\lesssim 0.5$ 
and starts to increase at $0.5< \mu\lesssim 0.6$.
Obviously there should be certain systematic errors 
associated with the $\alpha\to 0$ extrapolations,
in particular for $\mu=0.7$, where we cannot use the data points
at small $\alpha$.
Note, however, that the $\alpha$-dependence of the baryon number density 
shown in figure \ref{n small mu=all}
changes qualitatively between $\mu = 0.6$ and $\mu = 0.7$.
Since the extrapolations take this $\alpha$-dependence into account,
we consider that the rapid increase of the 
baryon number density at $\mu \sim 0.7$ is robust.

In figure \ref{n small mu=all r}, the squares
represent the results 
obtained by the RHMC simulation of the phase-quenched model.
In this case, we find that the baryon number density 
becomes nonzero at $\mu\sim 0.4$ and grows gradually 
with increasing $\mu$.
Thus we find that 
the onset of the baryon number density in the CLM occurs
at larger $\mu$
than in the phase-quenched model.
This is consistent with the theoretically expected behavior
known as the Silver Blaze phenomenon \cite{Cohen:2003kd}:
at zero temperature, the onset of the baryon number density 
in full QCD is expected to occur at $\mu=m_{N}/3$, 
while that in the phase-quenched model occurs 
at $\mu=m_{\pi}/2(<m_{N}/3)$, where $m_N$ is the nucleon mass
and $m_{\pi}$ is the pion mass. 
Obviously, the complex phase of 
the fermion determinant should play a crucial role 
in realizing the zero baryon number density 
within $m_{\pi}/2<\mu<m_{N}/3$ in full QCD.
The delayed onset of the baryon number density 
seen in our results of the CLM is a clear sign of the
Silver Blaze phenomenon, which
strongly suggests that the complex phase of the fermion determinant
is treated correctly in the CLM.
Let us also emphasize that we are able to obtain results
with the present setup up to $\mu/T = 0.7 \times 8 = 5.6$,
where the sign problem is extremely severe
and conventional lattice QCD simulations have never been successful.

\section{Summary and discussions}\label{sec-4}
In this work we showed for the first time
that it is possible to study QCD at high density and low temperature
by the CLM.
We have performed lattice QCD simulations with four-flavor 
staggered fermion on a $4^3\times8$ lattice and calculated 
the baryon number density
as a function of the quark chemical potential 
within $0.4\leq\mu\leq0.7$.
We made use of the gauge cooling 
and the deformation technique
for the excursion problem and the singular-drift problem, respectively.
The deformation technique was shown to be useful 
also in stabilizing the simulations. 
We checked the reliability of the obtained results 
by the probability distribution of the drift term.
The results for the original theory was obtained 
by extrapolating the reliable results to the zero deformation parameter.
The obtained results showed
a sharp onset of the baryon number density 
at $\mu\sim 0.6$. 

We also find that 
the onset of the baryon number density in the CLM 
occurs at larger $\mu$ than
in the phase-quenched model,
which is a clear sign of the Silver Blaze phenomenon.
In order to confirm this phenomenon,
we clearly need to increase the lattice size.
We have already started simulations
with a $8^3\times 16$ lattice \cite{Ito:ongoing}.
In this case, it turned out that correct results can be obtained
\emph{without deformation and extrapolations}
even in the region where the quark number density is not very small.
In that region, we obtained preliminary results showing
that the quark number density increases rapidly twice,
which may be interpreted as
the transition to the nuclear matter phase 
and that to the quark matter phase.
We hope to report on these results 
as well as the confirmation of the Silver Blaze phenomenon 
in the near future.


\section*{Acknowledgement}
K.~N.\ and J.~N.\ were 
supported in part by Grant-in-Aid for Scientific Research 
(No.~26800154 and 16H03988, respectively) 
from Japan Society for the Promotion of Science. 
S.~S.\ was supported by the MEXT-Supported Program 
for the Strategic Research Foundation at Private Universities 
``Topological Science'' (Grant No.~S1511006).

\bibliography{Lattice2017_148_SHIMASAKI_arxiv}

\end{document}